# Living with technology*

Eric Monteiro,

Dept. of computer and information science, Norwegian Univ. of Science and Technology

(NTNU), and Dept. of Informatics, Univ. of Oslo, (ericm@ifi.ntnu.no)

## 1. Introduction

Our use of IT has, broadly speaking, until recently evolved around work-oriented tasks (Ehn 1989; Greenbaum and Kyng 1991; Kyng and Mathiassen 1997). This pattern of use may crudely be characterized as:

- purposeful, work-oriented;
- performed by workers;
- delineated from other work tasks;
- restricted to working hours;

This is - or, indeed, has been – changing. Our present and future use of IT (or ICT or what have you) will clearly not be restricted along the same dimensions as those listed above. IT has moved beyond the work-place and into leisure, entertainment, games and our homes. It has in short moved into our everyday life. My retired father and oldest daughter use it. Use is no longer delineated but seeps into and get intervowen with a number of my activities: while paying for my groceries, planning this summer's vacation and so forth.

My aim here is not to plunge deeper into speculative manifests about what the future will bring. The point is simply to underline the fairly obvious observation that our use of IT is changing. Changing, I would hold, so much that our traditional notion of "use" of IT should be re-examined. It carries too much of a tool connotation, the image of a competent User mastering her Tool – even after modifying the image through an appeal to a more phenomenological understanding of a tool (Ehn 1989; Greenbaum and Kyng 1991; Suchman 1987). Instead of "using" technology, we should look at ways to explore how we live with technology.

Consider an example. An ATM, the automatic check-in teller at airports and electronic commerce on the net are basically the "same" in the sense that they are artefacts that allow a service (cash, travel or purchase). Despite the similarity, the "use" of this technology varies a lot, arguably more than is reasonably to ascribe to the differences in the artefacts themselves. We all use ATMs, some use check-in tellers and a few purchase on the net. The key reason, I suggest, why these differences prevail is that they are in quite different stages of being intervowen into the social fabric of life. How should this process be conceived of – we do not simply "use" these artefacts? How does it take place, what drives it, how important is your perception of what others do, how does trust get established, how many fellow passengers need to queue up in front of you before you inject your own ticket? These kinds of questions indicate the short-comings of the traditional notion of "using" technology.

---

* I am grateful for valuable comments on earlier drafts by Ole Hanseth, Pål Sørgaard and SJIS reviewers.

This has immediate implications for design as I see it. Being trapped in the traditional vocabulary of the "use" of technology is not merely a luxury. It is not that case that we, IS designers, can go on talking about the "use" of technology and leave a more fine-grained analysis to, say, social scientists with their queer disposition for details. On the contrary, I believe it is essential and high time for us to develop further our conceptualization of the "use" of IT in order to do better design.

In this brief comment, I will exemplify such an exploration by pointing out a few candidates for relevant, alternative notions and discuss their background and contents. Finally, I address the question of implications for IS research stemming from a re-conceptualization of what it means to introduce and "use" (old terminology is hard to avoid...) IT.

## 2. Consuming, taming or domesticating technology

The overall ambition for much of IS research for quite some time has been to develop concepts, theories and research methodologies which lend themselves to help unpacking how information systems come into being together with their subsequent "biography". The traditional separation into phases – development, introduction, use and diffusion – has serious short-comings which reduce its relevance. There are two major problems. It is based on a dogmatic and too clear distinction between the technical and the non-technical and a rather simplistic grasp of the unfolding dynamics. Both of these are serious flaws that IS research at the closing of the second millennium needs to move beyond.

Given that the very "use" of information systems is changing more or less along the lines indicated further above and that our traditional, phase-oriented conceptualization has weaknesses, where should we look for alternatives? Needless to say, there are numerous ones, and most scholars (not excluding myself) have their own pet candidates (see, for instance, (Ciborra 1996) and the notions of care and hospitality). Here, I point out two strands of research that so far have received very little attention in our community but which I suggest is relevant.

If our use of information systems is transforming over time, beyond the traditional, work-oriented tasks, where should we look for inspiration or analogous experience? The idea is fairly straightforward: explore a new phenomenon (new "use" of IT) by drawing on earlier, relevant experience. Quite often, I believe, the most fruitful way to explore a new phenomenon is by emphasizing its continuity with others rather than immediately yield to the urge to stress the discontinuity, to plunge into the new.

What, then, could possibly count as relevant experience? I suggest that the following two sources ought to be included on such a list:

- studies of mundane, *everyday technology* like washing machines, phone answering machines and television because they emphasize the new role for IT where we "mingle" around it without much ado;
- *historical studies* of how major technologies like telephone and electricity gradually got intervowen with and *embedded* into social life because they underscore the essential "socialization" of new technology into our lives;

Let me briefly illustrate them in turn.

**Domestication**

The motivation for looking at how we operate, perceive of and mingle with everyday technology is two-fold. First and most immediately, an interesting aspect of our new "use" of IT is the way it rapidly becomes natural, mundane or trivialized (Lie and Sørensen 1998).

Secondly, the way everyday technology is embedded underscore the need to "see the social and symbolic as well as material objects" (Silverstone and Hirsch 1992, p. 2). A focus on the everyday world encourages a richer and fuller grasp of human action – underscoring symbols, rituals, identity construction and values – than traditional, more or less functional explanations of purposeful, work-oriented tasks (Lie and Sørensen 1997; Silverstone and Hirsch 1992)

There are several, proposed metaphors for conceptualizing this including: "appropriation", "domestication", "consumption" and "taming" of technology (Lie and Sørensen 1998; Mackay 1997; Silverstone and Hirsch 1992). To focus on one, domestication is intended to emphasize the naturalization process, the way we cultivate and discipline artefacts when weaving them into the domestic sphere. It simultaneously underlines how "the wider world of work, leisure, and shopping are defined ... [and] are expressed in the specific and various cosmologies and rituals that define, or fail to define, the household's integrity as a social and cultural unit" (Silverstone and Hirsch 1992, p. 18). More precisely, the process of domestication may be seen to comprise of (ibid., p. 21):

- appropriation, where the access to the artefact is defined;
- objectification, through which various classificatory principles identify perceptions and claims for the status of the artefact;
- incorporation, the routinization or embedding into social patterns;
- conversion, the presentation of the constructed artefact to the world outside the domestic sphere, the household's contribution to the currencies of meanings;

All this is fine but what about the design of IT? For an increasing number of information systems, I would hold, concern about domestication is very much at the heart of design. Consider electronic purchasing on the net. Why is it – really – that actual "use" fails to live up to expectations and what can designers like ourselves do about it? For sure, there are numerous reasons, but certainly more than the narrow focus on security mechanisms. To illustrate an application of the notion of domestication, it would encourage a scrutiny of appropriation (access points that tie in with related tasks such as grocery orderings on the refrigerator front), objectification (articulate a set of scenarios connected to roles such as house wife, the business woman, the student, etc.), incorporation (alignment with institutionalized practices of shopping, of car driving, of organizing home life) and conversion (do people want to present and exhibit themselves through net shopping?).

**Social meanings of new technology: history of domestication**

The driving force behind looking at historical processes of domestication of technology is to develop a firmer grasp of the unfolding dynamics: how is it that new, flashy technology gets transformed into mundane, invisible technology? A telling example is Nye's (1990) description of how electricity seeped into the social fabric of a typical, small town in the United States at the turn of the century.

Again, electricity certainly did not get "introduced" and "used" in any straightforward sense. It constantly transformed itself – and social order – as it gradually seeped into the pores of modern, urban life.

To illustrate, the domestication of electricity in Nye's description extends well beyond simplistic dynamics of the development-introduction-use kind. Nye describes an evolving, dynamic and reflexive process whereby electricity get shaped by, but subsequently shapes, urban life. This unfolds through stumbling and improvised, yet comprehensive, exploration of the new technology. This process is not merely one of fitting the potential with the needs. Electricity feeds into the construction of modern, urban life. In doing so, it taps into deep sentiments and desires. For instance, the early "use" of electricity was lightning. Lightning, though, not of homes but of spectacle, highly profiled, public buildings with iconic or mythical status, for instance the New York Stock Exchange, La Scala opera and Niagara falls. This implied that their "social meaning resided not only in skillful electrical engineering but in the public perception of the new technologies as spectacle" (ibid., p. 58) which ultimately meant that electrification contributed to the construction of national identity as it "was becoming an essential part of experiencing both natural and natural symbols" (ibid., p. 61). Hence, it contributed to the "technological sublime" by simultaneously endowing awe and reverence (ibid., p. 59).

## 3. Relevance and implications for IS research

What possible reason can there be for IS researchers to bother about seemingly irrelevant issues like telephone answering machines and a hundred year old story about electricity? Only one, I believe, but an important one.

Much of our effort go into advancing our understanding about how the purely technical mesh with organizational, political and economical issues. This is an ongoing struggle where we slowly are making progress (Kling 1999; Walsham 1997). Still, we obviously have a long way to go. We are able to demonstrate how IT is socially constructed and exhibits a duality (Orlikowski 1992) or inscribes behaviour (Hanseth and Monteiro 1997). The problem, however, is that analysises like these tend not to pay justice to the richer specter of human life in the sense of incorporating identity, the home sphere and leisure. This implies that a conceptualization like actor-network theory (ANT), among the most developed frameworks on the market for studying the socio-technical I would argue, also has problems with accounting for behaviour that depart so radically from the (more or less) functional, work-oriented projects emphasized up till now. This was perhaps reasonable as long as IT was used in connection with work. As this is changing – IT seeps and gets embedded into most aspects of everyday life – we need to equip our studies with notions that cater for more of the symbolic aspects of modern life. To do so lies not on the margins, but would boost our abilities to design modern IT and address problems like it would take for, say, mobile IT to be widespread "used", that is, domesticated.